\documentstyle[12pt]{article}

\begin{document} 
\vskip 2cm
\begin{center}
{\large Influence of the additional second neighbor hopping on the spin
response in the $t$-$J$ model}\\
\vskip 1cm
Xianglin Ke and Feng Yuan \\
Department of Physics, Beijing Normal University, Beijing
100875, China \\
\vskip 1cm
Shiping Feng\\
$^{*}$Department of Physics, Beijing Normal University, Beijing
100875, China and \\
National Laboratory of Superconductivity, Academia Sinica, Beijing
100080, China and\\
Institute of Theoretical Physics, Academia Sinica, Beijing 100080,
China\\
\end{center}
\vskip 2cm

The influence of the additional second neighbor hopping $t'$ on the
spin response of the $t$-$J$ model in the underdoped and optimally
doped regimes is studied within the fermion-spin theory. Although
the additional second neighbor hopping $t'$ is systematically
accompanied with the reduction of the dynamical spin structure
factor and susceptibility, the qualitative behavior of the dynamical
spin structure factor and susceptibility of the $t$-$t'$-$J$ model
is the same as in the case of $t$-$J$ model. The integrated dynamical
spin structure factor spectrum is almost $t'$ independent, and the
integrated dynamical spin susceptibility still shows the particularly
universal behavior as
$I(\omega,T)\propto {\rm arctan}[a_{1}\omega/T+a_{3}(\omega/T)^{3}]$.

\leftline{71.27.+a, 74.72.-h, 76.60.-k}

\newpage

Following the initial discovery of the antiferromagnetic (AF) spin
fluctuation in copper oxide materials \cite{n1}, extensive
experimental and theoretical studies have been carried out in order
to clarify the relationship between the AF spin fluctuation and
superconductivity \cite{n2,n3,n4}. The single common feature of
copper oxide materials is the two-dimensional (2D) ${\rm CuO_{2}}$
plane \cite{n2,n10}, and it seems evident that the exotic behaviors
are dominated by this plane. It has been shown from the experiments
that the anomalous magnetic properties in copper oxide materials
mainly depend on the extent of dopings, and the regimes have been
classified into the undoped, the underdoped, the optimally doped,
and the overdoped, respectively \cite{n2,n3,n4}. The undoped copper
oxide materials are insulating systems \cite{n2}, and well understood
in terms of the 2D antiferromagnet with an AF long-range-order
(AFLRO) \cite{n3,n4}. This AFLRO is reduced dramatically with dopings
\cite{n5,n6}, and vanishes around the doping $\delta=5\%$. But a
series of inelastic neutron scattering measurements show that the AF
short-range spin fluctuation in copper oxide materials persists in
the underdoped and optimally doped regimes \cite{n7,n8,n9}. It is
widely believed that the same correlations that lead to the
insulating AF state at small doping, also lead to the
superconductivity in the underdoped and optimally doped regimes
\cite{n4,n10}. Since the copper oxide superconductors are doped Mott
insulators, many authors \cite{n11,n12} have suggested that the
essential physics of these materials can be effectively
described by the 2D $t$-$J$ model acting on the space with no doubly
occupied sites, where $t$ is the nearest neighbor hopping matrix
element, and $J$ is the nearest neighbor magnetic exchange
interaction. This model has been used to study the spin dynamics of
copper oxide materials in the underdoped and optimally doped regimes,
and the results obtained \cite{n4,n13,n14,n15} from the analytical
methods and numerical simulations are in qualitative agreement with
the experiments \cite{n7,n8,n9}.

However, the recent angle-resolved photoemission spectroscopy
measurements \cite{n16} on copper oxide materials show that
although the highest energy filled electron band is well described
by the $t$-$J$ model in the direction between the $(0,0)$ point and
the $(\pi,\pi)$ point in the momentum space, but both the
experimental data near $(\pi,0)$ point and overall dispersion may
be properly accounted by generalizing the $t$-$J$ model to include
the second- and third-nearest neighbors hopping terms $t'$ and
$t''$. These photoemission results also show that the electron band
width is reduced from the several eV expected from the band theory
to of order $J$, which indicates that the coupling of the electron
to the AF background plays an essential role in the electronic
structure \cite{n16}. On the other hand, the short-range AF spin
correlation in the underdoped and optimally doped regimes is
responsible for the nuclear magnetic resonance (NMR), nuclear
quadrupole resonance (NQR), and especially for the temperature
dependence of the spin-lattice relaxation rate \cite{n3}. It is
believed that both experiments from the angle-resolved photoemission
spectroscopy and neutron scattering measurements produce interesting
data that introduce important constraints on the microscopic models
and theories. In this case, a natural question is what is the effect
of these additional hoppings on the spin dynamics of the $t$-$J$
model. In this paper, we study this issue within $t$-$t'$-$J$ model.
Our results show that although the additional second neighbor
hopping $t'$ is systematically accompanied with the reduction of the
dynamical spin structure factor and susceptibility in the underdoped
and optimally doped regimes, the qualitative behavior of the
dynamical spin structure factor and susceptibility is the same as in
the case of the $t$-$J$ model \cite{n13,n14,n15}. The integrated
dynamical spin structure factor of the $t$-$t'$-$J$ model is almost
$t'$ independent, and the integrated dynamical spin susceptibility
still shows the particularly universal behavior as $I(\omega,T)
\propto {\rm arctan}[a_{1}\omega/T+a_{3}(\omega/T)^{3}]$.

We start from the 2D $t$-$t'$-$J$ model which can be written as,
\begin{eqnarray}
H=-t\sum_{i\hat{\eta}\sigma}C^{\dagger}_{i\sigma}C_{i+\hat{\eta}
\sigma}+t'\sum_{i\hat{\tau}\sigma}C^{\dagger}_{i\sigma}
C_{i+\hat{\tau}\sigma}+\mu\sum_{i\sigma}C^{\dagger}_{i\sigma}
C_{i\sigma}+J\sum_{i\hat{\eta}}{\bf S}_{i}\cdot
{\bf S}_{i+\hat{\eta}},
\end{eqnarray}
where $\hat{\eta}=\pm \hat{x},\pm\hat{y}$, $\hat{\tau}=\pm \hat{x}
\pm\hat{y}$, $C^{\dagger}_{i\sigma}$ ($C_{i\sigma}$) are the
electron creation (annihilation) operators, ${\bf S}_{i}=
C^{\dagger}_{i}{\bf \sigma}C_{i}/2$ are spin operators with $\sigma
=(\sigma_{x},\sigma_{y}, \sigma_{z})$ as the Pauli matrices, and
$\mu$ is the chemical potential. The $t$-$t'$-$J$ model (1) is
supplemented by the on-site local constraint $\sum_{\sigma}
C^{\dagger}_{i\sigma}C_{i\sigma}\leq 1$, ${\it i.e.}$, there are no
doubly occupied sites. The $t$-$J$ model was originally introduced
\cite{n11} as an effective Hamiltonian of the large-$U$ Hubbard
model, where the on-site Coulomb repulsion $U$ is very large as
compared with the electron hopping energy $t$, which leads to that
electrons become strongly correlated to avoid double occupancy.
Furthermore, many authors \cite{n12,n17} derived the $t$-$J$ model
or $t$-$t'$-$J$ model from a multiband large-$U$ Hubbard model
described with 2D ${\rm CuO_{2}}$ plane. Therefore the strong
electron correlation in the $t$-$J$ or $t$-$t'$-$J$ model manifests
itself by the electron single occupancy on-site local constraint.
This on-site local constraint can be treated exactly in analytical
calculations within the fermion-spin theory \cite{n18}, $C_{i\uparrow}
=h^{\dagger}_{i}S^{-}_{i}$ and $C_{i\downarrow}=h^{\dagger}_{i}
S^{+}_{i}$, where the spinless fermion operator $h_{i}$ keeps track
of the charge (holon), while the pseudospin operator $S_{i}$ keeps
track of the spin (spinon), then the fermion-spin theory naturally
incorporates the physics of the charge-spin separation. In the
fermion-spin representation, the $t$-$t'$-$J$ model can be expressed
\cite{n18} as,
\begin{eqnarray}
H &=& t\sum_{i\hat{\eta}}h^{\dagger}_{i+\hat{\eta}}h_{i}(S^{+}_{i}
S^{-}_{i+\hat{\eta}}+S^{-}_{i}S^{+}_{i+\hat{\eta}}) -
t'\sum_{i\hat{\tau}}h^{\dagger}_{i+\hat{\tau}}h_{i}(S^{+}_{i}
S^{-}_{i+\hat{\tau}}+S^{-}_{i}S^{+}_{i+\hat{\tau}}) \nonumber \\
&+& \mu \sum_{i}h^{\dagger}_{i}h_{i} + J_{eff}\sum_{i\hat{\eta}}
({\bf S}_{i}\cdot {\bf S}_{i+\hat{\eta}}),
\end{eqnarray}
where $J_{eff}=J[(1-\delta)^{2}-\phi^{2}_{1}]$, the holon
particle-hole order parameter $\phi_{1}=\langle h^{\dagger}_{i}
h_{i+\hat{\eta}}\rangle$, and $S^{+}_{i}$ and $S^{-}_{i}$ are the
pseudospin raising and lowering operators, respectively. As a
consequence, the kinetic part in the $t$-$t'$-$J$ model has been
expressed as the holon-spinon interactions in the fermion-spin
representation, which dominates the physics in the underdoped and
optimally doped regimes as in the case of the $t$-$J$ model
\cite{n14}.

Within the framework of the charge-spin separation, it has been
shown \cite{n20} that the charge dynamics can be discussed based on
the combination rule from spinons and holons, but no composition
law is required for discussing the spin dynamics, since the spin
fluctuation couples only to spinons, but the strongly correlation
between holons and spinons is considered through the holon's order
parameters entering in the spinon propagator. In this case, the
spin dynamics of the $t$-$J$ model in the underdoped and optimally
doped regimes has been discussed \cite{n14} within the fermion-spin
theory by considering spinon fluctuations around the mean-field
solution, where the spinon part is treated by the loop expansion to
the second-order. Following their discussions \cite{n14}, we can
obtain the dynamical spin structure factor and susceptibility in the
present $t$-$t'$-$J$ model as,
\begin{eqnarray}
S(k,\omega)&=&{\rm Re}\int^{\infty}_{0}dt e^{i\omega(t-t')}D(k,t-t')
=2[1+n_{B}(\omega)]{\rm Im}D(k,\omega),~~~~~\\
\chi''(k,\omega)&=&(1-e^{-\beta\omega})S(k,\omega)=2{\rm Im}
D(k,\omega),
\end{eqnarray}
respectively, where the full spinon Green's function,
$D^{-1}(k,\omega)=D^{(0)-1}(k,\omega)-\Sigma_{s}^{(2)}(k,\omega)$,
with the mean-field spinon Green's function, $D^{(0)-1}(k,\omega)=
(\omega^{2}-\omega^{2}_{k})/B_{k}$, and the second-order spinon
self-energy from the holon pair bubble,
\begin{eqnarray}
\Sigma_{s}^{(2)}(k,\omega)&=&\left ({Z\over N}\right )^{2}\sum_{pp'}
\gamma_{12}^{2}(k,p,p'){B_{k+p}\over 2\omega_{k+p}}\left ({F_{1}
(k,p,p')\over \omega+\xi_{p+p'}-\xi_{p'}-\omega_{k+p}} \right .
\nonumber \\
&-&\left. {F_{2}(k,p,p')\over \omega+\xi_{k+p'}-\xi_{p'}+
\omega_{k+p}}\right ),
\end{eqnarray}
where $\gamma_{12}(k,p,p')=t(\gamma_{k+p+p'}+\gamma_{k-p})-t'
(\gamma'_{k+p+p'}+\gamma'_{k-p})$, $\gamma_{{\bf k}}=(1/Z)
\sum_{\hat{\eta}}e^{i{\bf k}\cdot\hat{\eta}}$, $\gamma'_{{\bf k}}=
(1/Z)\sum_{\hat{\tau}}e^{i{\bf k}\cdot\hat{\tau}}$, $Z$ is the number
of the nearest neighbor or second-nearest neighbor sites, $B_{k}=
\Delta_{1}[2\chi^{z}_{1}(\epsilon\gamma_{k}-1)+\chi_{1}(\gamma_{k}-
\epsilon)]-\Delta_{2}(2\chi^{z}_{2}\gamma'_{k}-\chi_{2})$,
$\Delta_{1}=2ZJ_{eff}$, $\Delta_{2}=4Z\phi_{2}t'$, $\epsilon=1+2t
\phi_{1}/J_{eff}$, $F_{1}(k,p,p')=n_{F}(\xi_{p+p'})[1-n_{F}(\xi_{p'})]
+[1+n_{B}(\omega_{k+p})][n_{F}(\xi_{p'})-n_{F}(\xi_{p+p'})]$, $F_{2}
(k,p,p')=n_{F}(\xi_{p+p'})[1-n_{F}(\xi_{p})]-n_{B}(\omega_{k+p})[n_{F}
(\xi_{p'})-n_{F}(\xi_{p+p'})]$, $n_{F}(\xi_{k})$ and $n_{B}
(\omega_{k})$ are the fermion and boson distribution functions,
respectively, the mean-field holon spectrum $\xi_{k}=2Zt\chi_{1}
\gamma_{k}-2Zt'\chi_{2}\gamma'_{k}-\mu$, and the mean-field spinon
spectrum,
\begin{eqnarray}
\omega^{2}_{k}&=&\Delta_{1}^{2}\left ([\alpha C^{z}_{1}+{1\over 4Z}(1-
\alpha)-\alpha\epsilon\chi^{z}_{1}\gamma_{k}-{1\over 2Z}\alpha\epsilon
\chi_{1}](1-\epsilon\gamma_{k}) \right. \nonumber \\
&+&\left. {1\over 2}\epsilon[\alpha C_{1}+{1\over 2Z}(1-\alpha)-\alpha
\chi_{1}\gamma_{k}-{1\over 2}\alpha\chi^{z}_{1}](\epsilon-\gamma_{k})
\right ) \nonumber \\
&+&\Delta_{2}^{2}\left ([\alpha\chi^{z}_{2}\gamma'_{k}-{3\over 2Z}
\alpha\chi_{2}]\gamma'_{k}+{1\over 2}[\alpha C_{2}+{1\over 2Z}(1-
\alpha)-{1\over 2}\alpha\chi^{z}_{2}]\right ) \nonumber\\
&+&\Delta_{1}\Delta_{2}\left (\alpha\chi^{z}_{1}\gamma'_{k}(1-
\epsilon\gamma_{k})+{\alpha\over 2}(\chi_{1}\gamma'_{k}-C_{3})(\epsilon
-\gamma_{k})+\alpha\gamma'_{k}(C^{z}_{3}-\epsilon\chi^{z}_{2}
\gamma_{k})\right. \nonumber \\
&-&\left. {1\over 2}\alpha\epsilon(C_{3}-\chi_{2}\gamma_{k})\right ),
\end{eqnarray}
with the spinon correlation functions
$\chi_{1}=\langle S_{i}^{+}S_{i+\hat{\eta}}^{-}\rangle$,
$\chi_{2}=\langle S_{i}^{+}S_{i+\hat{\tau}}^{-}\rangle$,
$\chi^{z}_{1}=\langle S_{i}^{z}S_{i+\hat{\eta}}^{z}\rangle$,
$\chi^{z}_{2}=\langle S_{i}^{z}S_{i+\hat{\tau}}^{z}\rangle$,
$C_{1}=(1/Z)\sum_{\hat{\eta'}}\langle S_{i+\hat{\eta}}^{+}
S_{i+\hat{\eta'}}^{-}\rangle$,
$C^{z}_{1}=(1/Z)\sum_{\hat{\eta'}}\langle S_{i+\hat{\eta}}^{z}
S_{i+\hat{\eta'}}^{z}\rangle$,
$C_{2}=(1/Z)\sum_{\hat{\tau'}}\langle S_{i+\hat{\tau}}^{+}
S_{i+\hat{\tau'}}^{-}\rangle$,
$C_{3}=(1/Z)\sum_{\hat{\tau}}\langle S_{i+\hat{\eta}}^{+}
S_{i+\hat{\tau}}^{-}\rangle$,
$C^{z}_{3}=(1/Z)\sum_{\hat{\tau}}\langle S_{i+\hat{\eta}}^{z}
S_{i+\hat{\tau}}^{z}\rangle$, and the holon particle-hole order
parameter $\phi_{2}=\langle h^{\dagger}_{i}h_{i+\hat{\tau}}\rangle$.
In order not to violate the sum rule of the correlation function
$\langle S^{+}_{i}S^{-}_{i}\rangle=1/2$ in the case without AFLRO,
the important decoupling parameter $\alpha$ has been introduced in
the mean-field calculation, which can be regarded as the vertex
correction \cite{n19}.

For small dopings, the spin fluctuation scattering remains
commensurate at the AF wave vector $Q=(\pi,\pi)$ position
\cite{n2,n3}. With increasing dopings, there is a
commensurate-incommensurate transition in the spin fluctuation
geometry, and the incommensurate scattering in the underdoped and
optimally doped regimes corresponds to four 2D rods at
$(\pi\pm 2\pi\delta_{d},\pi)$ and $(\pi, \pi\pm 2\pi\delta_{d})$
with $\delta_{d}$ is the deviation of the peak position from the AF
wave vector $Q$ position \cite{n2,n3}. This incommensurate
scattering is the main new feature that appears into the
superconducting phase of copper oxide materials \cite{n2}. In this
paper, we are interested in the influence of the additional second
neighbor hopping on the spin dynamics of the $t$-$J$ model. To make
the discussion simpler, we only study the spin response of the
$t$-$t'$-$J$ model near the AF wave vector $Q$. We have performed a
numerical calculation for the dynamical spin structure factor
$S(Q,\omega)$ and dynamical spin susceptibility $\chi''(Q,\omega)$
of the $t$-$t'$-$J$ model, and the results of the $S(Q,\omega)$ and
$\chi''(Q,\omega)$ spectra at the doping (a) $\delta=0.06$ and (b)
$\delta=0.12$ with the temperature $T=0.2J$ for the parameters
$t/J$=2.5, $t'/J$=0.3 (solid line), and $t'/J$=0.5 (dashed line)
are plotted in Fig. 1 and Fig. 2, respectively. For comparison, the
corresponding results \cite{n14} of the $t$-$J$ model (dash-dotted
line) are also shown in Fig. 1 and Fig. 2, respectively. From Fig. 1
and Fig. 2, we find that although the additional second neighbor
hopping $t'$ is systematically accompanied with a clear reduction of
the dynamical spin structure factor and susceptibility in the
underdoped and optimally doped regimes, the qualitative behavior of
the dynamical spin structure factor and susceptibility in the
$t$-$t'$-$J$ model is the same as in the case of the $t$-$J$ model
\cite{n14}. The spin structure factor spectrum is separated into low-
and high-frequency parts, respectively, but the high-frequency part
is suppressed in the susceptibility, then the low-frequency peak
dominates the dynamical spin susceptibility, the neutron-scattering,
and NMR processes, which is consistent with the experiments
\cite{n2,n7,n8}.

One of the most important features of the spin dynamics in copper
oxide materials is the universal behavior of the integrated
dynamical spin response \cite{n2,n9}. This universal behavior is
very significant because of its relation to many other normal state
properties of copper oxide materials. The integrated dynamical spin
response is manifested by the integrated dynamical spin structure
factor and integrated dynamical spin susceptibility, and can be
expressed as,
\begin{eqnarray}
\bar{S} (\omega)&=&S_{L}(\omega)+S_{L}(-\omega)=(1+e^{-\beta
\omega})S_{L}(\omega)=(1+e^{-\beta \omega}){1\over N}\sum_{k}
S(k,\omega),~~~~\\
I(\omega, T)&=&{1\over N}\sum_{k}\chi^{\prime\prime}(k,\omega),
\end{eqnarray}
respectively. The numerical results of the integrated dynamical spin
structure factor $\bar{S}(\omega)$ at the doping (a) $\delta=0.06$
and (b) $\delta=0.12$ with the temperature $T=0.2J$ for the
parameters $t/J$=2.5, $t'/J$=0.3 (solid line), and $t'/J$=0.5
(dashed line) are shown in Fig. 3. The dash-dotted line is
the corresponding result \cite{n14} of the $t$-$J$ model. These
results indicate that the integrated spin structure factor of the
$t$-$t'$-$J$ model is almost $t'$ independent in the underdoped and
optimally doped regimes. Moreover, $\bar{S}(\omega)$ is decreased
with increasing energies for $\omega <0.5t$, and constant for
$\omega >0.5t$. In correspondence with the integrated dynamical spin
structure factor, the numerical results of the integrated dynamical
spin susceptibility at the doping $\delta=0.12$ with the temperature
$T=0.2J$ for the parameters $t/J$=2.5, $t'/J$=0.3 (solid line), and
$t'/J$=0.5 (dashed line) are shown in Fig. 4. The dash-dotted line
is the corresponding result \cite{n14} of the $t$-$J$ model. Our
results show that the integrated susceptibility increases with
increasing $\omega/T$ for $\omega/T <1$, and then is almost constant
for $\omega/T >1$. Although the value of the integrated dynamical
spin susceptibility of the $t$-$t'$-$J$ model is weak $t'$ dependent,
but the shape still is particularly universal, and can be scaled
approximately as $I(\omega,T)=b_{1}{\rm arctan}[a_{1}\omega/T+a_{3}
(\omega/T)^{3}]$ as in the case of the $t$-$J$ model \cite{n14}.
These results are in very good agreement with the experiments
\cite{n9}.

The $t$-$J$ model is characterized by a competition between the
kinetic energy ($t$) and magnetic energy ($J$). The magnetic energy
$J$ favors the magnetic order for spins, while the kinetic energy
$t$ favors delocalization of holes and tends to destroy the magnetic
order. Only in this sense, the additional second neighbor hopping
$t'$ in the $t$-$J$ model is equivalent to increase the kinetic energy,
and its influence on the spin dynamics of the $t$-$J$ model may be
similar to the effect of dopings. On the other hand, the scattering
of spinons dominates the spin dynamics, and the qualitative behavior
of the spin dynamics in the $t$-$J$ model is not changed dramatically
with dopings in the underdoped and optimally doped regimes \cite{n14}.
These are why at least for small values of $t'$ the qualitative
behavior of the spin dynamics in the $t$-$t'$-$J$ model is the same
as these obtained from the $t$-$J$ model. Since the scattering of
holons dominates the charge dynamics \cite{n21}, and some qualitative
physical properties of the charge dynamics are changed dramatically
with dopings in the $t$-$J$ model \cite{n21}, then it is possible
that the additional second neighbor hopping $t'$ may affect the
qualitative behavior of the charge dynamics of the $t$-$J$ model.

In summary, we have discussed the influence of the additional second
neighbor hopping $t'$ on the spin response in the $t$-$J$ model in
the underdoped and optimally doped regimes within the fermion-spin
theory. Our results show that although the additional second neighbor
hopping $t'$ is systematically accompanied with the reduction of the
dynamical spin structure factor and susceptibility, the qualitative
behavior of the dynamical spin structure factor and susceptibility
of the $t$-$t'$-$J$ model is the same as in the case of $t$-$J$ model.
The integrated spin structure factor spectrum is almost $t'$
independent, and the integrated dynamical spin susceptibility still
shows the particularly universal behavior as
$I(\omega,T)\propto {\rm arctan}[a_{1}\omega/T+a_{3}(\omega/T)^{3}]$.

\vskip 2cm
\centerline{Acknowledgements}

The authors would like to thank Professor H.Q. Lin, Professor C.D.
Gong and Professor Z.X. Zhao for helpful discussions. This work was
supported by the National Natural Science Foundation of China under
Grant No. 19774014.

\newpage
\centerline{Figure Captions}

Figure 1. The dynamical structure factor $S(Q,\omega)$ at the doping
(a) $\delta=0.06$ and (b) $\delta=0.12$ with the temperature $T=0.2J$
for the parameters $t/J$=2.5, $t'/J$=0.3 (solid line), and $t'/J$=0.5
(dashed line). The dash-dotted line is the corresponding result of
the $t$-$J$ model.

Figure 2. The dynamical susceptibility $\chi''(Q,\omega)$ at the doping
(a) $\delta=0.06$ and (b) $\delta=0.12$ with the temperature $T=0.2J$
for the parameters $t/J$=2.5, $t'/J$=0.3 (solid line), and $t'/J$=0.5
(dashed line). The dash-dotted line is the corresponding result of
the $t$-$J$ model.

Figure 3. The integrated dynamical structure factor $\bar{S}(\omega)$ at
the doping (a) $\delta=0.06$ and (b) $\delta=0.12$ with the temperature
$T=0.2J$ for the parameters $t/J$=2.5, $t'/J$=0.3 (solid line), and
$t'/J$=0.5 (dashed line). The dash-dotted line is the corresponding
result of the $t$-$J$ model.

Figure 4. The integrate dynamical susceptibility $I(\omega)$ at the
doping $\delta=0.12$ with the temperature $T=0.2J$ for the parameters
$t/J$=2.5, $t'/J$=0.3 (solid line), and $t'/J$=0.5 (dashed line). The
dash-dotted line is the corresponding result of the $t$-$J$ model.

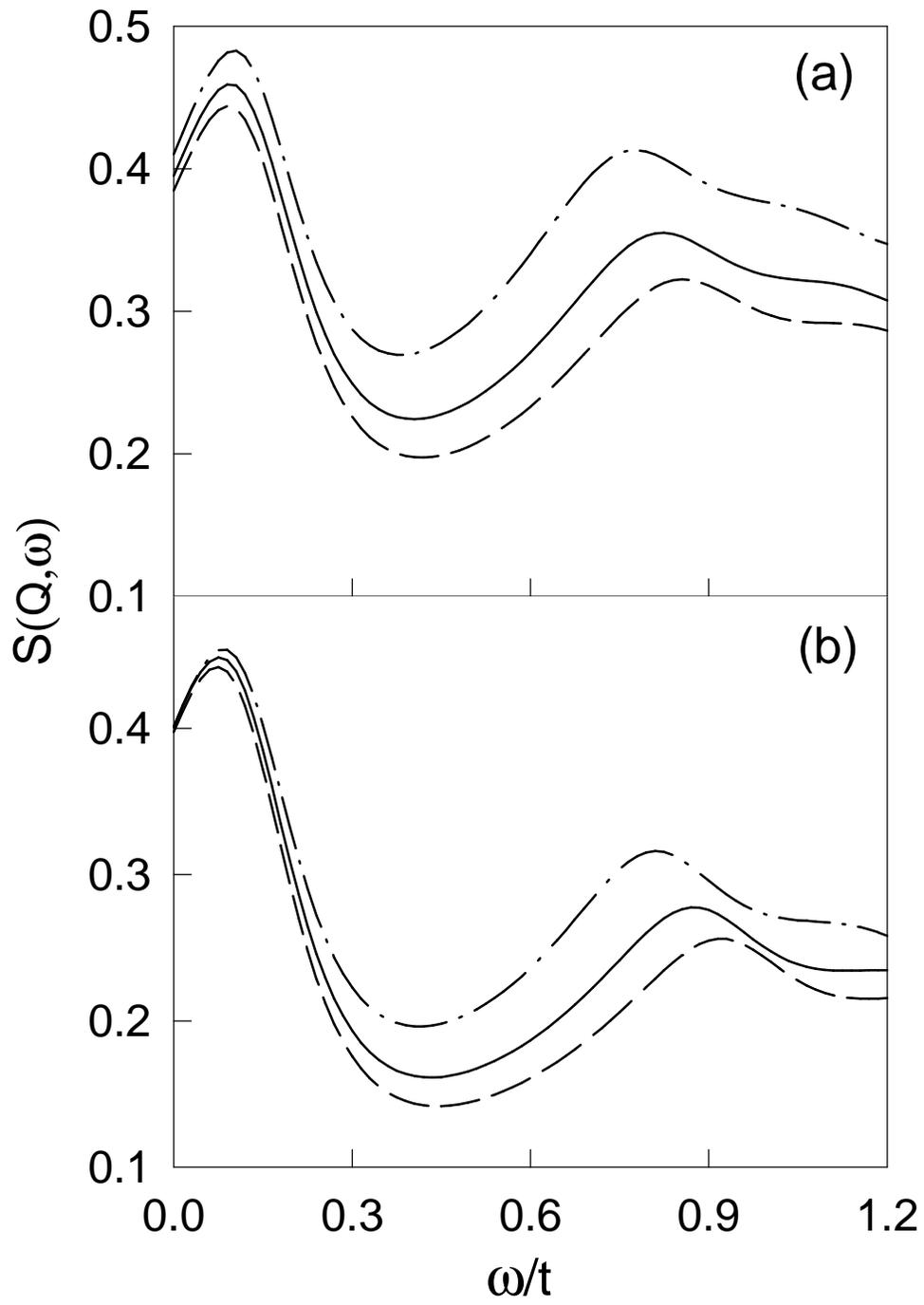
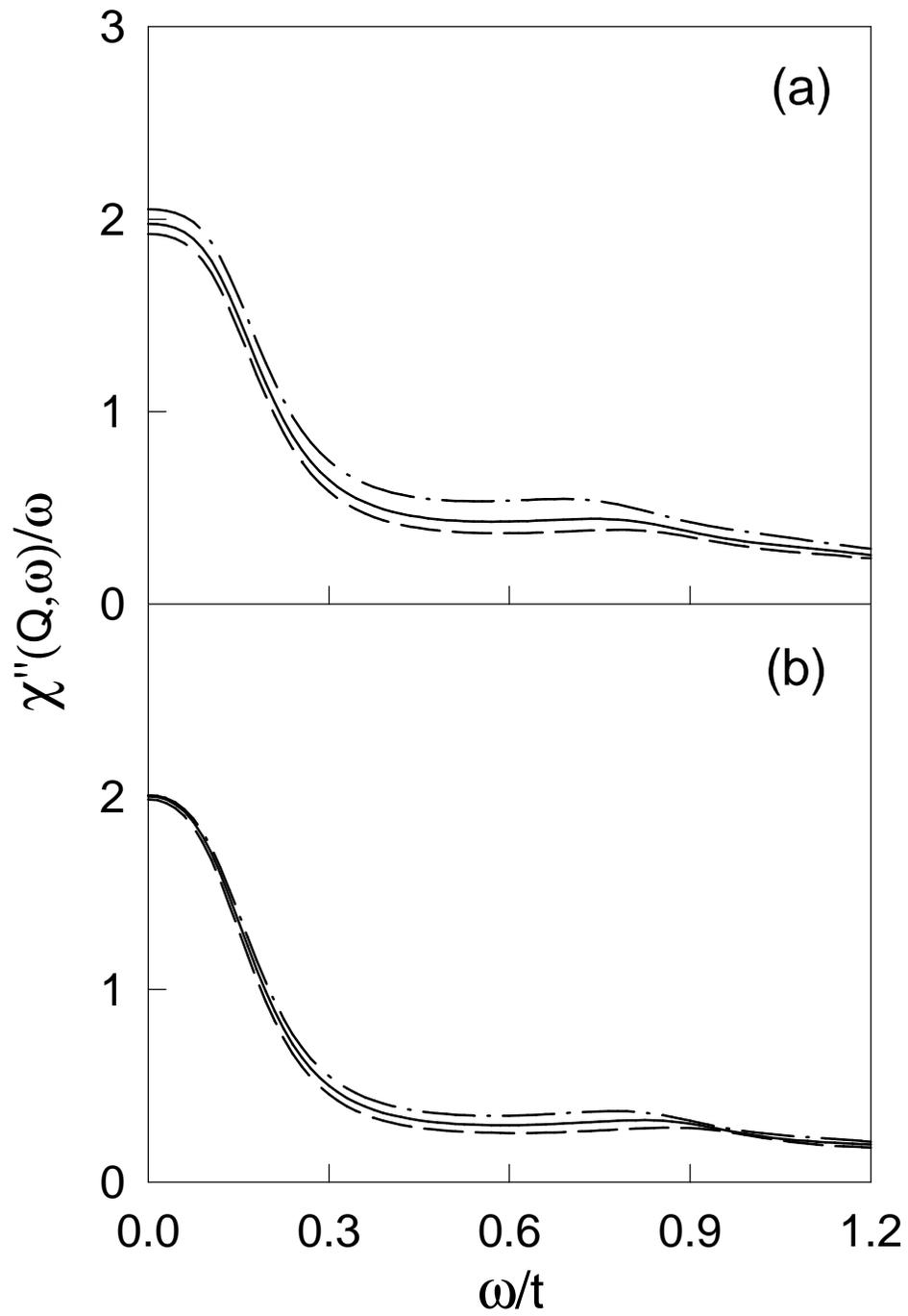
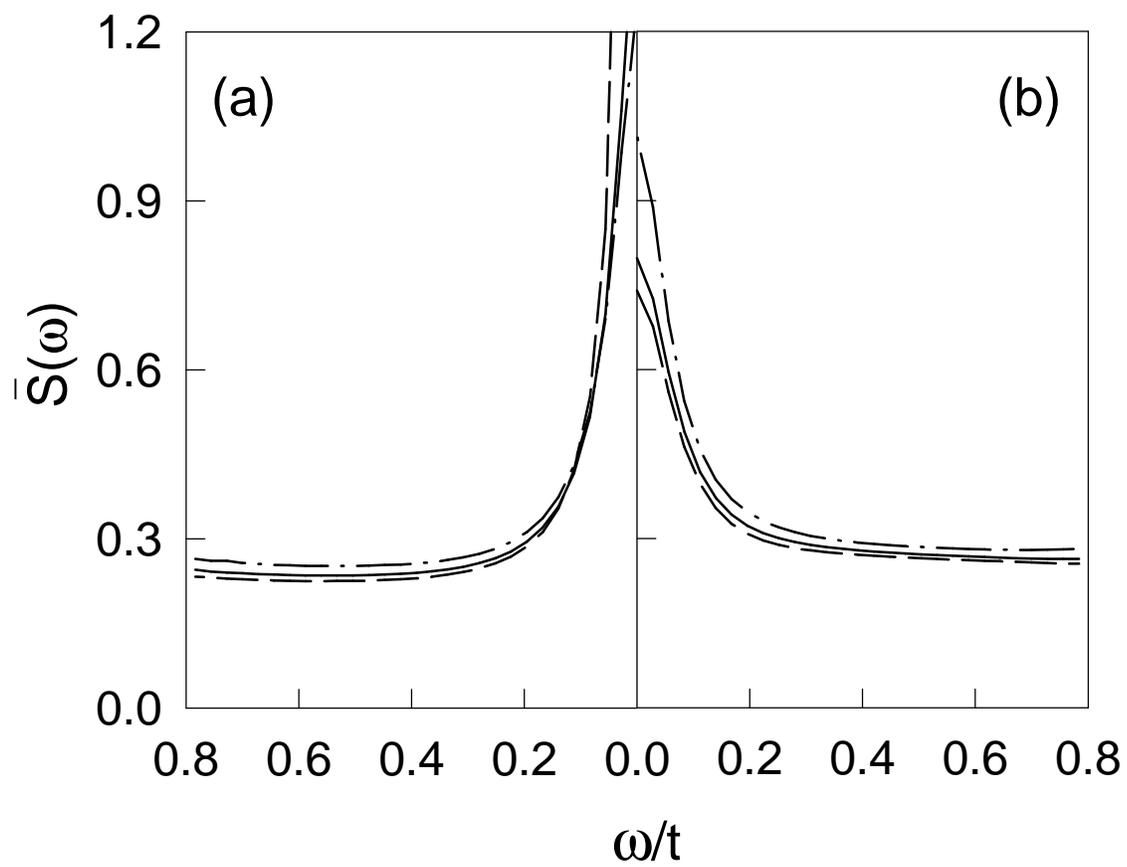
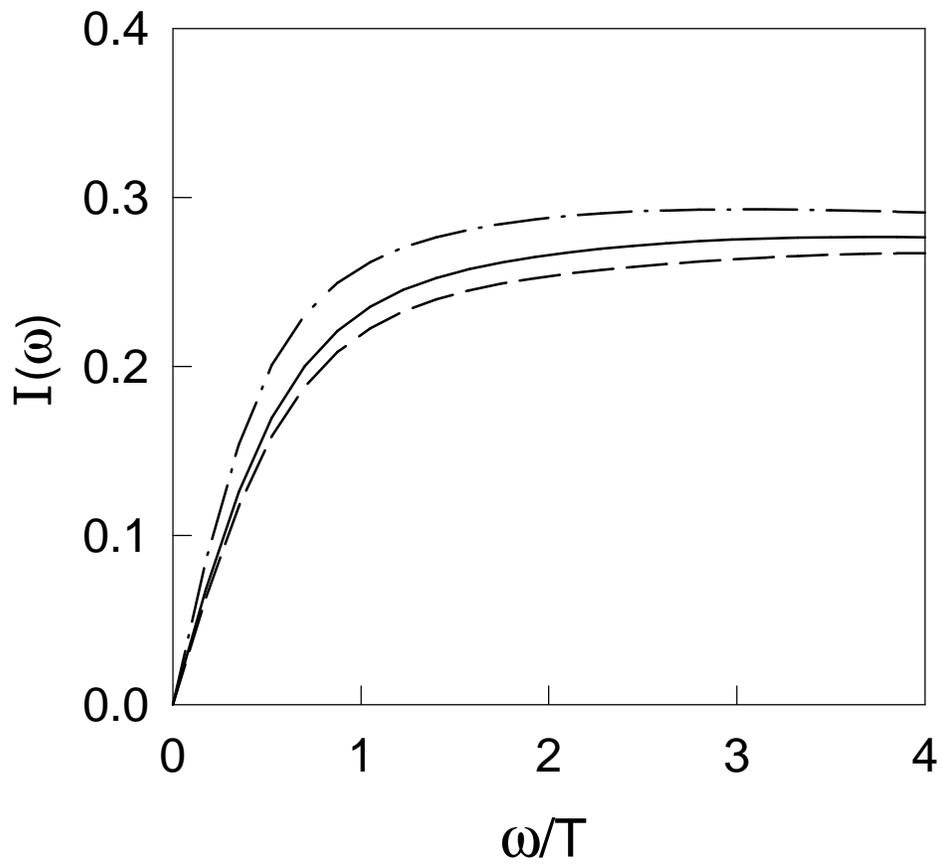
\end{document}